# Time-Rate-Transformation framework for targeted assembly of short-range attractive colloidal suspensions

*Safa Jamali[1], Robert C. Armstrong[2] and Gareth H. McKinley[3]*

[1] *Dept. of Mechanical and Industrial Engineering, Northeastern University, Boston MA 02115*
[2] *Dept. of Chemical Engineering, Massachusetts Institute of Technology, Cambridge MA 02139*
[3] *Dept. of Mechanical Engineering, Massachusetts Institute of Technology, Cambridge MA 02139*

*Abstract-* The aggregation of attractive colloids has been extensively studied from both theoretical and experimental perspectives as the fraction of solid particles is changed, and the range, type and strength of attractive or repulsive forces between particles varies. The resulting gels consisting of disordered assemblies of attractive colloidal particles, have also been investigated with regards to percolation, phase separation, and the mechanical characteristics of the resulting fractal networks. Despite tremendous progress in our understanding of the gelation process, and the exploration of different routes for arresting the dynamics of attractive colloids, the complex interplay between convective transport processes and many-body effects in such systems has limited our ability to drive the system towards a specific configuration. Here we study a model attractive colloidal system over a wide range of particle characteristics and flow conditions undergoing aggregation far from equilibrium. The complex multiscale dynamics of the system can be understood using a Time-Rate-Transformation diagram adapted from understanding of materials processing in block copolymers, supercooled liquids and much stiffer glassy metals to direct targeted assembly of attractive colloidal particles.

*Main-* The physical and mechanical properties of soft condensed matter generally, and colloidal systems specifically, are directly governed by their underlying microstructure. The diversity of different microstructures obtained are the results of: (i) interactions between the primary particles that constitute the material [1, 2], and (ii) the interplay between the complex many-body effects that arise from these interactions and the processing conditions under which the material system is fabricated [3, 4]. Attractive Brownian particles assemble into colloidal gels, glasses and crystals depending on the volume fraction of colloidal particles and their inter-particle interactions [5]; each state representing an ensemble microstructure that particles collectively form [4]. Although one can predict theoretically the most favorable global structure for the particle assembly, the interplay between dynamical transport processes and the growing structures, coupled with the collective hydrodynamic behavior of large particle clusters, makes it extremely challenging to control the final configuration of such structured materials. Since the collective energy landscape of the particle ensemble in a given condition exhibits a plethora of local minima that the particulate structures can explore, kinetically arrested colloidal gel states form, rather than the ordered crystalline microstructure associated with the deepest energy minimum [3, 6]. Colloidal gels typically exhibit arrested microstructures, in which the individual particles form partially-ordered space-spanning networks that determine the mechanical response and properties of the resulting soft material. Recent advances in imaging techniques, as well as computational studies of particle assembly under quiescent conditions, have brought invaluable insight into the microscopic physics and dynamics of these particulate structures [6-10]; however, the resulting colloidal assemblies evolve and dynamically respond to imposed boundary conditions, resulting in a thermokinematic memory of the flow history [11, 12], which leads to shear-induced structuration/rejuvenation [13-15], as well as rheological hysteresis and thixotropy [16, 17].

Recent experimental and computational studies have shown that the dynamical behavior of colloidal gels can be systematically understood based on the ratio of the magnitude of shear forces arising from flow to the strength of attraction between colloidal particles. This ratio is commonly referred to as a Mason number, $Mn \sim (\text{shear forces})/(\text{attractive forces}) = 6\pi\eta\dot{\gamma}a^3/\Gamma$ [12, 18-20], where $\eta$ is the suspending fluid viscosity, $a$ is the radius of a particle, $\dot{\gamma}$ is the imposed deformation rate, and $\Gamma$ is the strength of attraction between a pair of Brownian particles (typically expressed in $kT$ units). In similar approaches, others have proposed use of the term "Peclet number of depletion [18, 20] which is effectively a Mason number normalized or rescaled by the range of attraction [21]; however, since the Mason number is defined as the ratio of the *magnitude* of shear forces to attractive forces exerted on a particle, here we do not normalize the Mason number by the range of attraction. At small deformation rates, $Mn < 10^{-2}$, (where shear forces promoting bond breakage are significantly smaller than attractive forces promoting bond formation) the material behaves similarly to a viscoelastic solid and bulk deformation tests show that it exhibits a yield stress [22]. Above this critical stress, irreversible flow of the material begins, with progressive deviation from linear viscoelastic solid behavior (often characterized by a transport stress overshoot) and ultimately steady flow similar to a viscoplastic fluid [18, 23-26]. At sufficiently strong deformation rates, $Mn > 10^0$, the particulate network is effectively broken down into individual components and the system response to an imposed deformation is increasingly hard-sphere-like [27]. At intermediate rates [28], $10^{-2} \leq Mn \leq 10^0$, constant breakage and reformation of particle-particle bonds results in dynamic evolution of flow heterogeneities over a wide range of scales and the evolution of complex secondary structures that change the bulk stress response of the fluid as a whole [25, 28, 29]. Nonetheless, it should be noted that the use of the Mason number in scaling the rheological behavior of colloidal gels is limited to particles with short-range attraction. For longer-range potentials the shape of the energy landscape and the functional form of the potential needs to be taken into account. Thus, here and for the entirety of our results, we study behavior of colloidal systems with attractions acting within quarter of the particle radius in order to avoid potential-dependent structures. Furthermore one should note that use of Mason number as defined here can only be justified for spherical particles with symmetric interactions. In many practical applications, where the shear forces are exerted on non-spherical particles, or on particles with asymmetric interaction potentials, similar dimensionless groups will need to be defined accordingly.

Because of these complex dynamics, the resulting microstructures, as well as the final mechanical properties of a colloidal gel, strictly depend not only on the flow rates at which the material was processed, but also on the entire processing history [12]. Consequently, regardless of advances in particle chemistry, and physical understanding of the self-assembly phenomenon [30-33], systematic methods for fabrication of a colloidal network with specific or tailored final properties requires understanding of the non-equilibrium rheology of the self-assembling systems. The seminal work of Koumakis et al. [20] has demonstrated the role of processing history, by carefully examining the structure and rheology of shear-rejuvenated colloidal gels at different imposed deformation rates, and after cessation of flow. Nonetheless, a clear quantitative route to desired structure for a given system is lacking.

Lu et al. [6] prescribed a phase diagram of colloidal gelation, with clear description of the solid-like to liquid-like boundary, based on 3 different parameters: the volume fraction of solid particles, the interaction strength and the range of attraction in-between particles. Other theoretical/computational phase diagrams have been proposed based on similar parameters [3, 4, 34]. However, these phase maps are specified on the basis of our understanding of quiescent gelation and do not reflect the time-dependent response of the system to deformation history. For example, a specified volume fraction of particles with specific range and strength of attraction, well within the solid-like behavior region of the quiescent state diagram may show minimal to no elastic behavior in small amplitude oscillatory deformation depending on the history of the flow in which it was formed [35]. The present work develops a road map to understanding shear-mediated assembly of colloidal particles by considering the effect of flow history and outlines how to construct a dynamical phase map in which such microstructures can be tuned towards desirable mechanical properties. To do so, we perform numerical simulations incorporating many-body hydrodynamic interactions of attractive colloidal particles well within the gel-forming region of Lu's phase map for associating particles at $\phi = 0.15$ and study the effect of accumulated strain as well as the strength of the flow on the final microstructure and properties of the fluid.

There is a growing recognition that there are close analogies between the material response of stiff engineering materials such as metals and glasses and soft microstructured materials such as colloidal or protein gels [36, 37]. Here we extend this connection by adapting a well-known framework from material processing studies of complex engineering materials with a wide range of possible flow-induced microstructures known as the Time-Temperature-Transformation diagram [38-42]. This allows us to systematically unravel many of the complex features associated with shear-driven assembly of complex colloidal structures. Since flow strength can be characterized by the shear rate, or in dimensionless form by the Mason number, and the accumulated strain provides a convenient dimensionless measure of the time over which transport processes take place, we refer to this framework as a *Time-Rate-Transformation* diagram (or alternatively as a *Strain-Mason Number-Transformation* diagram). It should be noted that the Péclet number (i.e. the ratio of the shear forces to the thermal forces) is commonly used to describe the rheological behavior of hard-sphere suspensions; however, in these attractive colloidal systems we use the rescaling $Mn = Pe \times (kT/\Gamma)$ as the appropriate dimensionless group of choice.

We perform Dissipative Particle Dynamics [18] simulations (for implementation details see the SI) of attractive suspensions with solid volume fraction of $\phi = 0.15$ [25000 spherical colloidal particles and 200000 solvent particles], using a Morse potential to represent the inter-particle interaction as $U = -\Gamma(2e^{-\kappa h_{ij}} - e^{-2\kappa h_{ij}})$ where $h_{ij}$ is the surface-surface distance between two interacting particles. In this scheme, colloidal particles interact through Morse potential, as well as random, dissipative and lubrication forces, excluding traditional conservative forces in order to avoid discontinuities in particle interaction potentials (For details of simulation method refer to Supplementary Material and [12, 18]). Having the exact expression for the interaction potentials, one can accordingly calculate the effective range of attraction using the Noro-Frankel expression [43]; however, it has been shown that the stickiness parameter and the second virial component of all interaction potentials become virtually identical at short distances where the separation between two particles is smaller than 0.1-0.3$a$ [43, 44]. We show this quantitatively in Table S3 of the Supplementary Materials. In our simulations we study a wide range of attraction strengths $3kT \leq \Gamma \leq 50kT$, and three different ranges of attraction ($\kappa^{-1}/a = 0.1, 0.2$ and $0.25$), and keep the volume fraction constant at $\phi = 0.15$. Furthermore for all shear simulations a Lees-Edwards boundary condition is employed in order to impose deformation rates.

We have previously studied gelation under quiescent conditions for a range of volume fractions $0.1 \leq \phi \leq 0.3$, and measured the corresponding ensemble averaged coordination numbers, $Z_0$, as a suitable measure of the ensemble microstructure [12]. These gels are then subject to a steady simple shear flow of strength $\dot{\gamma}_0$, and the ensemble-average coordination numbers at long times, $Z(\dot{\gamma}_0; t \to \infty)$, were again measured for different applied shear rates. When plotted against the Mason number, the normalized coordination number [scaled with the value measured at the end of gelation under quiescent conditions, $Z_{eq}$] confirms our previous findings of three distinct regimes in these systems (Fig. 1). This behavior can be well described by a scaled expression of the form (equation 1):

$$\mathcal{Z} = \frac{Z(\dot{\gamma}_0; t \to \infty)}{Z_0} = \frac{Z_\infty}{Z_0} + \frac{d(1+\mathrm{Mn})}{1+b\mathrm{Mn}^c} \tag{1}$$

where $Z_\infty$ is the average coordination number of a hard-sphere suspension at a given shear rate ($\Gamma = 0kT$), $d = 1-(Z_\infty/Z_0)$ and the two fitting parameters are found to be, $b = 0.3$ and $c = 2$. The small fluctuations at low Mason number observed in fig.1 are associated with irreversible shear-aging of the colloidal gels under a slow, but non-zero, imposed deformation rate. In this regime, commonly referred to as "shear compaction", although the shear stresses imposed by the fluid flow are not sufficient to break the particulate network globally, they provide the particles near the surface of locally-glassy clusters [45] with additional energy required to occasionally break their bonding interactions and explore a more stable global state with progressively larger number of bonds [12, 18]. Recently, Whitaker et al. [45] through graph theory and structural analysis of the particle configurations showed that in gels formed under quiescent conditions as well as under flowing conditions, the elasticity of the gel arises from the packing of locally glassy clusters.

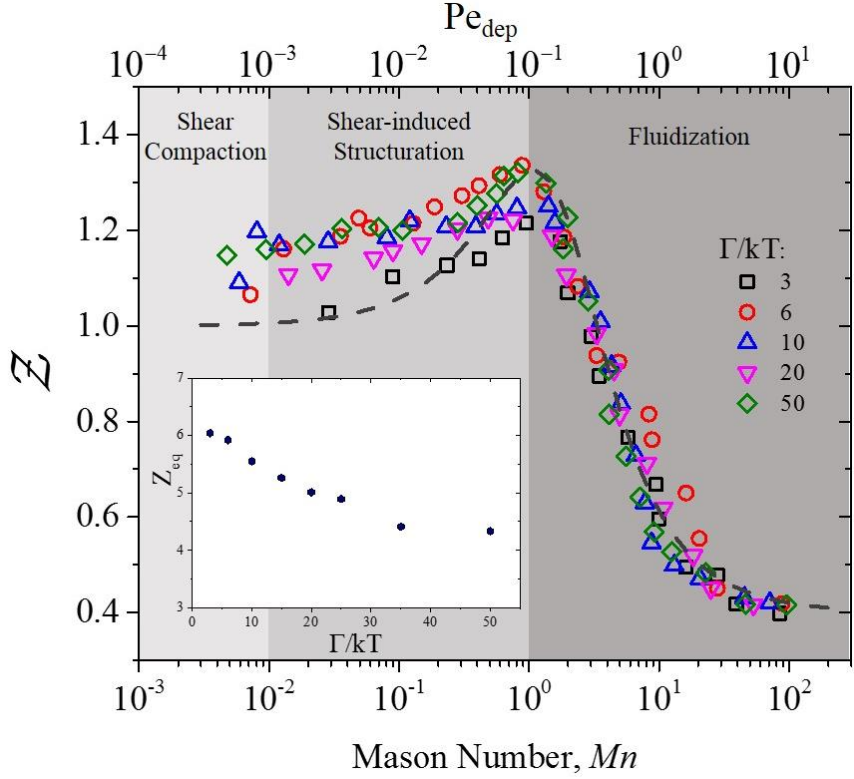

Figure 1. Evolution in the ensemble-averaged coordination number (scaled by its value at quiescent conditions), for a range of attraction strengths, in attractive colloidal systems as a function of applied Mason number.

Having the long-time rheology-microstructure relationship of these attractive colloids under flow at hand, we prescribe the following protocol for the Time-Rate-Transformation (TRT): all simulations are initiated with a random distribution of 25000 colloidal particles at $Mn=10$ to ensure complete rejuvenation of the structure [12], and then all simulations are brought to arrest, $\dot{\gamma}(t_q)=0$, at different quench times, $t_q$ with a linear decrease in shear rate over time of the form:

$$\dot{\gamma}(t) = \dot{\gamma}_0 (1 - t/t_q) \qquad (2)$$

Based on the results in fig. 1, and since at $Mn>1$ particle-particle bonds are effectively broken, time is initialized at the instance when the imposed deformation rate corresponds to $Mn=1$, i.e. $\dot{\gamma}_0 = \Gamma / 6\pi\eta_0 a^3$ (note that time is expressed in DPD units as explained in the SI).

The total accumulated strain, $\gamma_{max.} = \int \dot{\gamma}(t) dt = \dot{\gamma}_0 t_q / 2$ (i.e. the integrated area under the shear rate-time curve) is an initial obvious measure of the accumulated flow history; however, since three different ranges of attractions are investigated, a more universal choice is to incorporate the length scale of the attraction and we report results using a rescaled strain, $\gamma_n = (\kappa^{-1}/a)\gamma_{max.}$. Note that in our description, the Mason number only captures the ratio of the magnitude of shear forces to attractive forces. One can alternatively choose to represent the data against the "Peclet number of depletion, $Pe_{dep}$" as defined by Koumakis et al. [20], including the range of attraction in the dimensionless shear rate. However, using the range of attraction between the particles to rescale the effective accumulated strain, and the Mason number to represent the relative magnitude of the forces, the resulting dimensionless group is equivalent to using $Pe_{dep}$ and the total strain. Upon being brought to rest, another $10^3 \tau$ is allowed for the particulate structure to reach its steady state [where $\tau$ is the Brownian diffusive time of the colloid, equal to $a^2/D = 6\pi\eta a^3/kT$ with $D$ the diffusion coefficient of a single particle]. The rescaled time-time stress autocorrelation function, $C_{\alpha\beta}(t) = V \langle \sigma_{\alpha\beta}(t)\sigma_{\alpha\beta}(t+\Delta t) \rangle / kT$ (averaged over all three independent off-diagonal components of the total stress tensor, $\sigma_{\alpha\beta}$, where $V$ is the volume of the calculation cell), decays to zero for a viscous fluid, with the area under the autocorrelation curve being a measure of the zero shear viscosity of the system [46, 47]. One can decouple the contribution of different interactions to the total stress by individually investigating the interactions between particles; however, here the reported stress autocorrelation is calculated based on the total stress tensor and including all interactions (i.e. solvent-solvent, solvent-colloid, and colloid-colloid) in order to represent relevant experimentally measured bulk stresses. Each stress autocorrelation graph is averaged over 500 time origins ($t$), with 10 time-step intervals (each time step being $\Delta t = 0.0005 r_c (m/kT)^{0.5}$ where *m* is mass of individual DPD particles), resulting in the largest error bar being smaller than 1% of the calculated quantities. By contrast, for a yield stress fluid, at long times $C_{\alpha\beta}$ decays to a finite value [48] which is a measure of the residual stress, $\sigma_r$ in the system. In fig.2a we show the evolution of the stress autocorrelation: while the simple exponential decay of $C_{\alpha\beta}(t)$ for the hard sphere suspension indicates existence of a well-defined zero-shear viscosity [49], a non-zero terminal value of the same measure for the attractive particle suspension leads to a divergent area integral and prevents a zero-shear viscosity from being defined for these systems [22]. The oscillations in the stress autocorrelation function of the attractive systems correspond to nonlinear viscoelasticity and emergence of a yield stress, with the period of oscillation being proportional to the quench times imposed during the TRT protocol (see SI). Fig.2b shows the residual stress in the attractive suspensions for different attraction strengths and ranges, as a function of the total normalized accumulated strain under which they have been brought to rest. The solid line presented is fit to $\sigma_r = \sigma_{r,eq}/(1+\gamma_n)$, where $\sigma_{r,eq}$ is the residual stress of the attractive gel formed under quiescent conditions ($\dot{\gamma} = 0$). Normalizing the residual stress values of different gels by the corresponding values in the same gel formed at no-flow conditions, collapses all data, regardless of the strength and range of attraction, into a master curve. Evidently, the colloidal structures formed over longer quench times show increasingly fluid-like behavior (similar to flocculated clusters of particles) whereas the identical particle ensembles prepared in the same way, but quenched over shorter times, result in gels with increasingly solid-like properties.

Careful examination of the structures quenched over the shortest times ($\gamma_n \leq 2$), show slightly larger residual stresses than the ones formed at no-flow conditions. This arises from the annealing actions of small viscous shear stresses which allow particles to break out of their cages, and explore a lower energy state by forming more bonds, as evidenced by the small *Mn* limit of fig.1 in the shear compaction regime.

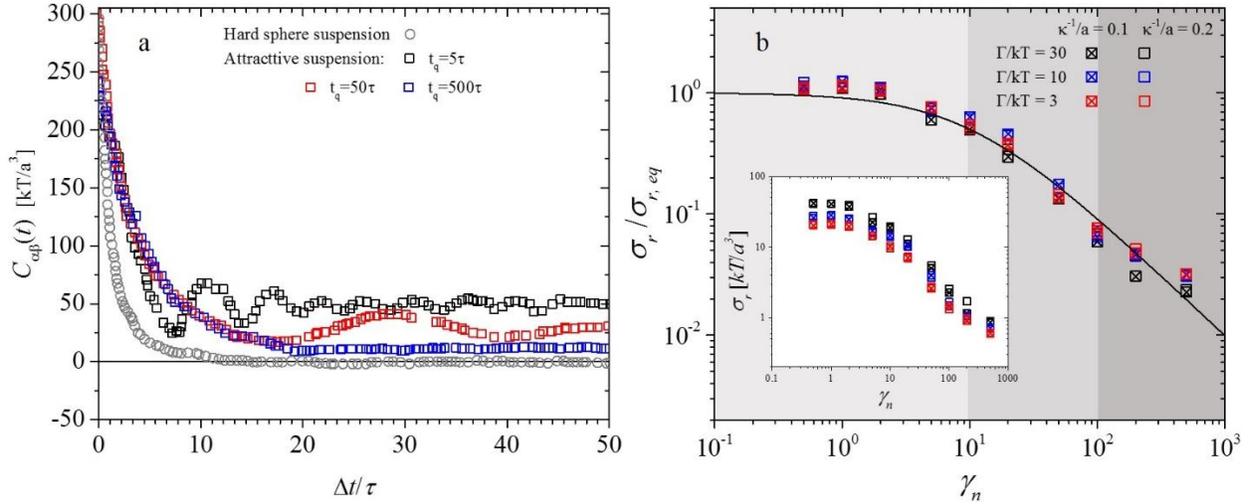

Figure 2. a) Stress autocorrelation function against time measured for hard sphere suspension quenched from $Pe = 1$ over $t_q = 10\tau$ and an attractive gel with $\kappa^{-1} = 0.1a$ and $\Gamma = 30k_BT$ quenched from $Mn = 1$, and b) Normalized residual stress against normalized accumulated strain for attractive systems under our TRT protocol, at different quench times. The line shows fit to $\sigma_r = \sigma_{r,eq} / (1 + \gamma_n)$.

In recent work, we have shown that the time- and rate-dependent rheology of these attractive colloidal systems can be parametrized through distinct signatures in micro- (average coordination number, <Z>), meso- (number density fluctuations, NDF), and macro- (shear stress) scale measures of a system under flow. Fig. 3a shows scaled values of the average coordination number, $\mathcal{Z}$, and number density fluctuations, $\mathcal{N}$, of the final structures, as a function of total accumulated strain for the same suspensions as in fig. 2b. The dashed lines represent corresponding values of the same quantities measured for the same system under quiescent gelation.

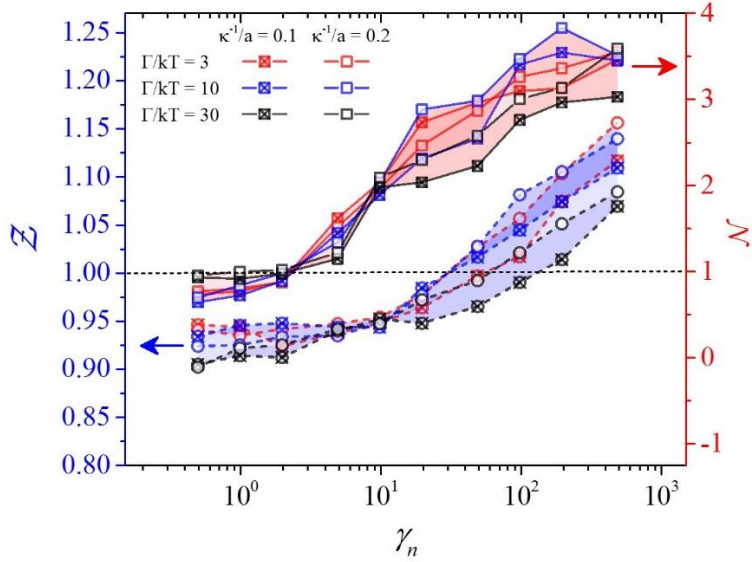

Figure 3. Ensemble-averaged coordination number (shaded in blue color, with square symbols and solid lines) and normalized number density fluctuations, $\mathcal{N}$ (shaded in red color with circular symbols and dashed lines) scaled by their values at quiescent conditions of gels brought to rest at different quench times. The dotted line represents the unity for both measures, corresponding to gel formed at rest.

Based on our previous work and the results in fig.2/3, we can define clear criteria for three distinct structural states formed under different conditions: (i) *strong gels* with $\sigma_r/\sigma_{r,eq} \approx O(1)$, $\mathcal{N} \leq 1$ and $\mathcal{Z} \leq 1$ where the system exhibits elastic solid-like behavior; (ii) *weak gels* corresponding to $0.05 \leq \sigma_r/\sigma_{r,eq} \leq 0.5$, $\mathcal{N} > 1$ and $\mathcal{Z} \leq 1$, where a percolated network of particles is present and thus the material exhibits a finite residual stress, consistent with the presence of a yield stress and a rheological behavior similar to a soft viscoelastic solid, but the yield stress is smaller than observed in gels formed isotropically and under no-flow conditions; and (iii) *flocculated fluids* with no clear indication of a yield stress where the system no longer exhibits gel-like behavior, a percolated network of particles is absent, and the mechanical properties are significantly weaker than the strong gels, $\sigma_r/\sigma_{r,eq} \leq 0.05$, $\mathcal{N} > 1$ and $\mathcal{Z} > 1$. The suspensions quenched over the shortest times and smallest strains, $\gamma_n < 10$, not only result in the strongest gels with the largest residual stress ($\sigma_r$), but they also have distinctly different micro- and meso-scale structures than the gels formed under quiescent conditions. Analysis of the fractal dimension of the gels using a box counting algorithm [12] shows that at intermediate shearing times, $10 < \gamma_n < 100$, the strong gels are denser with larger fractal dimensions ($D^f = 2.21$ for $\gamma_n = 1$ compared to $D^f = 1.76$ for $\gamma_n = 20$).

Using these criteria elucidated above, we can construct a phase map of the different regimes and the corresponding microstructures (Fig.4) using our direct numerical simulations of each regime. The distinctions between strong gel, weak gel and flocculated fluid states correspond to the material configurations at the end of the quench protocol, i.e. under no flow condition, and the shaded regions in this time-rate-transformations process (Fig. 4(top)) are pathways to those final structures. Finite Mason numbers indicate that the system is "fluidized" and exposed to an imposed deformation rate, and cannot therefore strictly be referred to as a gel. Fig. 4(bottom) summarizes the final states at the end of the TRT protocol over 285 simulation results for varying quench times and strengths of attraction between particles. Since the initial time required for calculation of strain is associated with the condition $Mn = 1$, (and thus varies for different applied shear rates and different values of $\Gamma$), a given value of of $\gamma_n$ corresponds to different quench times for different attraction strengths. In order to incorporate the effects of quench time and attraction strength we re-normalize the values of $\gamma_n$ using the characteristic thixotropic timescale (scaled by the diffusive timescale) for each attraction strength, $\tau_{thix.}(\Gamma)$. As we have shown in previous work this thixotropic timescale is a strong function of $\Gamma$, and varies as $\tau_{thix.}/\tau \sim (\Gamma/kT)^{-0.32}$ [see SI for additional details]. The resulting phase diagram clearly shows that, regardless of the particle attraction strength or range of attraction, for long quench times (corresponding to $\gamma_n . \tau_{thix.}/\tau \geq 400$) the particles effectively form phase-separated clusters that are not inter-connected. On the other hand, for very rapid quenches when the total shearing time for the particles to explore the distribution of local energy minima is short, $\gamma_n \cdot \tau_{thix.}/\tau \leq 40$, strong, disordered, fractal and percolated gels are formed. In the intermediate regime of $40 < \gamma_n \cdot \tau_{thix.}/\tau < 400$, although some connections between discrete particulate clusters are present, the meso-structure is heterogeneous with significantly larger number density fluctuations compared to quiescent gelation, and the resulting weak gel shows a much lower yield stress when a deformation is imposed.

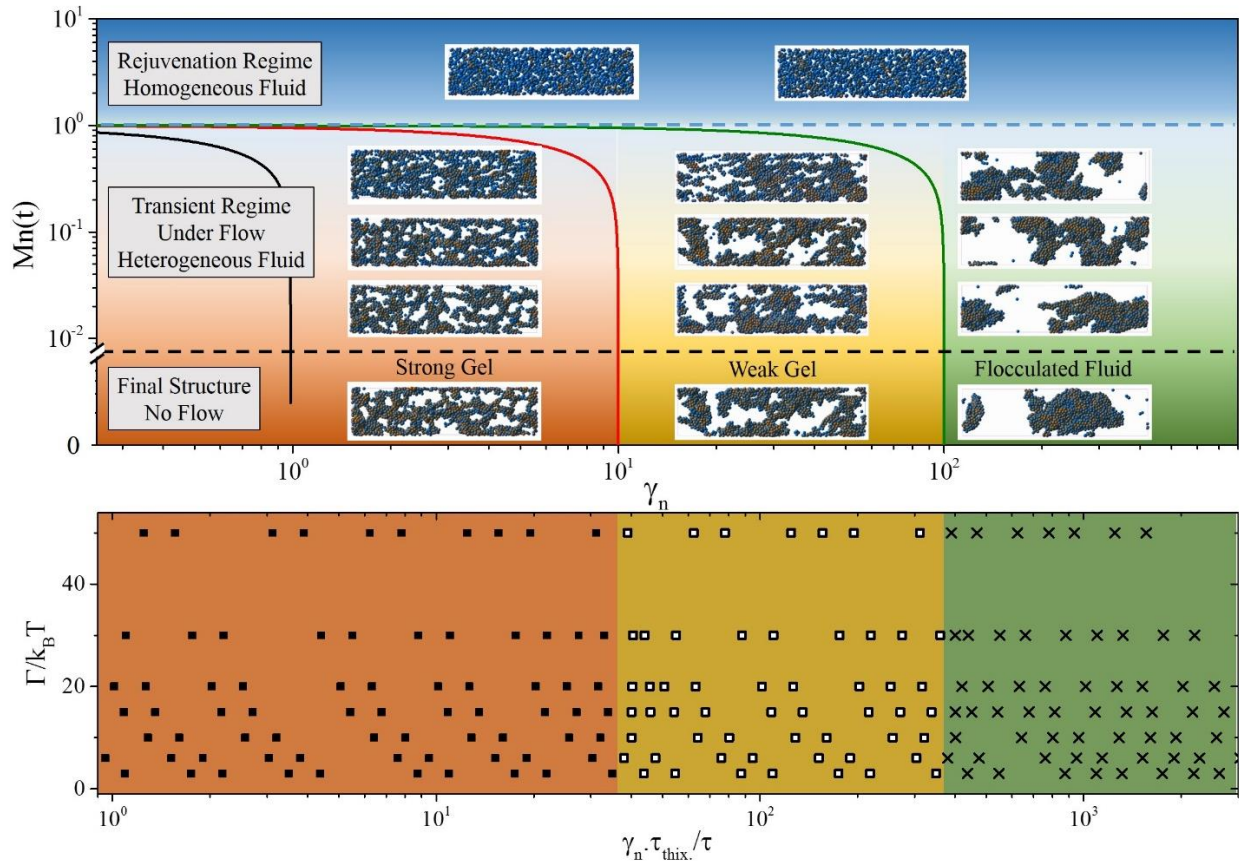

Figure 4. Strain-Mason number-Transformation phase diagram, with representative structure snapshots at each corresponding region for a system of Γ=10kT and $\kappa^{-1}$=0.1$a$. The top figure shows the Mason number and the normalized accumulated strains, with three solid lines showing linear shear rate decay with $t_q$ = 2, 20, and 200 for black, red and the green line respectively. The break in the ordinate axis is to distinguish between final structures at the end of flow protocol, and under no flow conditions. The lower figure shows the final state of the material (■ for strong gel, □ for weak gel, and × for flocculated fluid) as a function of an appropriately-rescaled strain accumulated during the quench process. This strain measure has been rescaled to account for the systematic variation in the relevant thixotropic time scale of each colloidal system as the attraction strengths between the particles are varied (see Fig. S3 for an un-rescaled representation).

These strain-Mason number state diagrams for attractive colloidal particles provide a systematic guide for predicting the type of structure/property that one can obtain for a given volume fraction, the strength and the range of attraction between the particles. This Time-Mason Number representation clearly shows that for these strictly time-dependent materials the final structure is not merely a function of the initial state variables but that a clear understanding of the underlying transport processes including the evolution in the imposed shear rate, the timescale of the deformation history and the relevant thixotropic timescale of the material are essential for predicting the final system state. Our results further confirm the role of shearing protocols experimentally investigated by Divoux *et al.* [16] and Helal *et al.* [35], in which the shear history is shown to control the macroscopic measures of the final structures, i.e. shear viscosity, loss and storage moduli, as well as electrical conductivity. Furthermore, fast confocal microscopy of sheared attractive colloids have also shown the role of prior shear history and the rate of applied deformation in controlling the microstructural evolutions of colloidal gels under flow [25, 27, 45]. We have shown that the final structures of a quiescent gel-forming system can vary from a dense fractal gel with a large yield stress, to disconnected fluid flocs. We have prescribed a Time-Rate-Transformation procedure, based on the Mason number which represents the (time-evolving) strength of the flow (compared to strength of attraction between the particles) and a normalized accumulated strain, $\gamma_n$, which incorporates both the range of attraction as well as the flow history. The resulting dynamical phase map provides alternative routes to shear-mediated design of attractive colloidal gels with desired structures. Although a linear quench profile (eq.2) was investigated in detail here, one can also design different routes through this strain-Mason number diagram (in the SI we also explore exponential profiles as well as step-wise decay of deformation rate with different waiting times at each rate) and the resulting state maps remain largely unchanged. Our findings, coupled with this simple dimensionless representation of the material history experienced by these thixotropic attractive particulate systems, enable us to design and control *a priori* colloidal assembly into different rheological and microstructural states.

**References**

1. Poon, W., *Colloids as Big Atoms.* Science, 2004. **304**(5672): p. 830.
2. Trappe, V., et al., *Jamming phase diagram for attractive particles.* Nature, 2001. **411**(6839): p. 772-775.
3. Zaccarelli, E., *Colloidal gels: equilibrium and non-equilibrium routes.* Journal of Physics: Condensed Matter, 2007. **19**(32): p. 323101.
4. Zaccarelli, E. and W.C.K. Poon, *Colloidal glasses and gels: The interplay of bonding and caging.* Proceedings of the National Academy of Sciences, 2009. **106**(36): p. 15203-15208.
5. Sciortino, F., *One liquid, two glasses.* Nature Materials, 2002. **1**(3): p. 145-146.
6. Lu, P.J., et al., *Gelation of particles with short-range attraction.* Nature, 2008. **453**(7194): p. 499-503.
7. Bergenholtz, J., W.C.K. Poon, and M. Fuchs, *Gelation in model colloid-polymer mixtures.* Langmuir, 2003. **19**(10): p. 4493-4503.
8. Furukawa, A. and H. Tanaka, *Key role of hydrodynamic interactions in colloidal gelation.* Physical Review Letters, 2010. **104**: p. 245702(1-4).
9. Helgeson, M.E., et al., *Homogeneous percolation versus arrested phase separation in attractively-driven nanoemulsion colloidal gels.* Soft Matter, 2014. **10**(17): p. 3122-3133.
10. Poon, W.C.K., A.D. Pirie, and P.N. Pusey, *Gelation in colloid-polymer mixtures.* Faraday Discussions, 1995. **101**(0): p. 65-76.
11. Geri, M., et al., *Thermokinematic memory and the thixotropic elasto-viscoplasticity of waxy crude oils.* Journal of Rheology, 2017. **61**(3): p. 427-454.
12. Jamali, S., G.H. McKinley, and R.C. Armstrong, *Microstructural rearrangements and their rheological implications in a model Thixotropic Elastoviscoplastic fluid.* Physical Review Letters, 2017. **118**(4): p. 048003.
13. Grenard, V., N. Taberlet, and S. Manneville, *Shear-induced structuration of confined carbon black gels: steady-state features of vorticity-aligned flocs.* Soft Matter, 2011. **7**(8): p. 3920-3928.
14. Gisler, T., R.C. Ball, and D.A. Weitz, *Strain Hardening of Fractal Colloidal Gels.* Physical Review Letters, 1999. **82**(5): p. 1064-1067.
15. Osuji, C.O. and D.A. Weitz, *Highly anisotropic vorticity aligned structures in a shear thickening attractive colloidal system.* Soft Matter, 2008. **4**(7): p. 1388-1392.
16. Divoux, T., V. Grenard, and S. Manneville, *Rheological hysteresis in soft glassy materials.* Physical Review Letters, 2013. **110**(1): p. 018304.
17. Larson, R.G., *Constitutive equations for thixotropic fluids.* Journal of Rheology, 2015. **59**(3): p. 595-611.
18. Boromand, A., S. Jamali, and J.M. Maia, *Structural fingerprints of yielding mechanisms in attractive colloidal gels.* Soft Matter, 2017. **13**(2): p. 458-473.
19. Scirocco, R., J. Vermant, and J. Mewis, *Effect of the viscoelasticity of the suspending fluid on structure formation in suspensions.* Journal of Non-Newtonian Fluid Mechanics, 2004. **117**(2): p. 183-192.

20. Koumakis, N., et al., *Tuning colloidal gels by shear.* Soft Matter, 2015. **11**(23): p. 4640-4648.
21. Varga, Z., et al., *Hydrodynamics control shear-induced pattern formation in attractive suspensions.* Proceedings of the National Academy of Sciences, 2019. **116**(25): p. 12193-12198.
22. Bonn, D., et al., *Yield stress materials in soft condensed matter.* Reviews of Modern Physics, 2017. **89**(3): p. 035005.
23. Colombo, J. and E. Del Gado, *Stress localization, stiffening, and yielding in a model colloidal gel.* Journal of Rheology, 2014. **58**(5): p. 1089-1116.
24. Gibaud, T., D. Frelat, and S. Manneville, *Heterogeneous yielding dynamics in a colloidal gel.* Soft Matter, 2010. **6**(15): p. 3482-3488.
25. Mohraz, A. and M.J. Solomon, *Orientation and rupture of fractal colloidal gels during start-up of steady shear flow.* Journal of Rheology (1978-present), 2005. **49**(3): p. 657-681.
26. Park, J.D. and K.H. Ahn, *Structural evolution of colloidal gels at intermediate volume fraction under start-up of shear flow.* Soft Matter, 2013. **9**(48): p. 11650-11662.
27. Hsiao, L.C., et al., *Role of isostaticity and load-bearing microstructure in the elasticity of yielded colloidal gels.* Proceedings of the National Academy of Sciences, 2012. **109**(40): p. 16029-16034.
28. Rajaram, B. and A. Mohraz, *Microstructural response of dilute colloidal gels to nonlinear shear deformation.* Soft Matter, 2010. **6**(10): p. 2246-2259.
29. Mohraz, A., E.R. Weeks, and J.A. Lewis, *Structure and dynamics of biphasic colloidal mixtures.* Physical Review E, 2008. **77**(6): p. 060403.
30. Colombo, J. and E. Del Gado, *Self-assembly and cooperative dynamics of a model colloidal gel network.* Soft Matter, 2014. **10**(22): p. 4003-4015.
31. Glotzer, S.C. and M.J. Solomon, *Anisotropy of building blocks and their assembly into complex structures.* Nature Materials, 2007. **6**: p. 557.
32. Sacanna, S., L. Rossi, and D.J. Pine, *Magnetic click colloidal assembly.* Journal of the American Chemical Society, 2012. **134**(14): p. 6112-6115.
33. Solomon, M.J., *Directions for targeted self-assembly of anisotropic colloids from statistical thermodynamics.* Current Opinion in Colloid & Interface Science, 2011. **16**(2): p. 158-167.
34. Varga, Z. and J. Swan, *Hydrodynamic interactions enhance gelation in dispersions of colloids with short-ranged attraction and long-ranged repulsion.* Soft Matter, 2016. **12**(36): p. 7670-7681.
35. Helal, A., T. Divoux, and G.H. McKinley, *Simultaneous rheoelectric measurements of strongly conductive complex fluids.* Physical Review Applied, 2016. **6**(6): p. 064004.
36. Keshavarz, B., et al., *Nonlinear viscoelasticity and generalized failure criterion for polymer gels.* ACS Macro Letters, 2017. **6**(7): p. 663-667.
37. Schmoller, K.M. and A.R. Bausch, *Similar nonlinear mechanical responses in hard and soft materials.* Nature Materials, 2013. **12**: p. 278.
38. Yamada, M., et al., *Interplay between Time-Temperature Transformation and the liquid-liquid phase transition in water.* Physical Review Letters, 2002. **88**(19): p. 195701.
39. Lee, S., C. Leighton, and F.S. Bates, *Sphericity and symmetry breaking in the formation of Frank–Kasper phases from one component materials.* Proceedings of the National Academy of Sciences, 2014. **111**(50): p. 17723-17731.

40. Gillard, T.M., S. Lee, and F.S. Bates, *Dodecagonal quasicrystalline order in a diblock copolymer melt.* Proceedings of the National Academy of Sciences, 2016. **113**(19): p. 5167.
41. Moore, E.B. and V. Molinero, *Structural transformation in supercooled water controls the crystallization rate of ice.* Nature, 2011. **479**: p. 506.
42. Peker, A. and W.L. Johnson, *Time-temperature-transformation diagram of a highly processable metallic glass.* Materials Science and Engineering: A, 1994. **179-180**: p. 173-175.
43. Noro, M.G. and D. Frenkel, *Extended corresponding-states behavior for particles with variable range attractions.* The Journal of Chemical Physics, 2000. **113**(8): p. 2941-2944.
44. Lekkerkerker, H.N. and R. Tuinier, *Colloids and the depletion interaction*. Vol. 833. 2011: Springer.
45. Whitaker, K.A., et al., *Colloidal gel elasticity arises from the packing of locally glassy clusters.* Nature Communications, 2019. **10**(1): p. 2237.
46. Kubo, R., *Statistical-mechanical theory of irreversible processes. I. General theory and simple applications to magnetic and conduction problems.* Journal of the Physical Societ of Japan, 1957. **12**(6): p. 570-586.
47. Boromand, A., S. Jamali, and J.M. Maia, *Viscosity measurement techniques in Dissipative Particle Dynamics.* Computer Physics Communications, 2015. **196**: p. 149-160.
48. Whittle, M. and E. Dickinson, *Brownian dynamics simulation of gelation in soft sphere systems with irreversible bond formation.* Molecular Physics, 1997. **90**(5): p. 739-758.
49. Lin, N.Y.C., M. Bierbaum, and I. Cohen, *Determining quiescent colloidal suspension viscosities using the Green-Kubo relation and image-based stress measurements.* Physical Review Letters, 2017. **119**(13): p. 138001.

Supplementary material for

# "Time-Rate-Transformation framework for targeted assembly of short-range attractive colloids"

*Safa Jamali, Robert C. Armstrong and Gareth H. McKinley*

**DPD and time-integration details**

The equation of motion for DPD can be written in the form of eq. S1, as the sum of pairwise interactions between two particles within a cut-off distance, $r_c$, of each other [1-3]. In classical DPD, the total force acting on a particle is calculated from three individual interactions: the random force, $\mathbf{F}_{ij}^R$, the dissipative force, $\mathbf{F}_{ij}^D$, and the conservative force, $\mathbf{F}_{ij}^C$ resulting in the force balance:

$$m_i \frac{d\mathbf{v}_i}{dt} = \sum \mathbf{F}_{ij}^R + \mathbf{F}_{ij}^D + \mathbf{F}_{ij}^C \tag{S1}$$

The first two terms on the right side of eq. S1 (random and dissipative interactions) form the thermostat and are essential for satisfaction of the fluctuation-dissipation theorem. The random force is a Brownian fluctuation source with a random function $\mathbf{F}_{ij}^C$, $\Theta_{ij}$ possessing a zero mean and a unit variance:

$$\mathbf{F}_{ij}^R = \sigma_{ij} \omega_{ij}(\mathbf{r}_{ij}) \frac{\Theta_{ij}}{\sqrt{\Delta t}} \mathbf{e}_{ij} \tag{S2}$$

Here $\Delta t$ is the time step used in the simulation, $\sigma_{ij}$ is the strength of the thermal fluctuations in the system, and $\mathbf{e}_{ij}$ is the unit vector $\mathbf{e}_{ij} = r_{ij}/|r_{ij}|$.

The dissipative force (eq. S3) is a heat sink acting against the relative motion of particles ($\mathbf{v}_{ij} = \mathbf{v}_i - \mathbf{v}_j$). In eq. S3, $\gamma_{ij}$ is the strength of dissipation which is coupled with the thermal noise, $\sigma_{ij}$, to control the temperature in the system. Together the parameters in Eqs (S2) and (S3) define the dimensionless DPD temperature as $kT = \sigma_{ij}^2 / 2\gamma_{ij}$.

$$\mathbf{F}_{ij}^D = -\gamma_{ij}(\omega_{ij}(\mathbf{r}_{ij}))^2 (\mathbf{v}_{ij} \cdot \mathbf{e}_{ij}) \mathbf{e}_{ij} \tag{S3}$$

The last interaction represents the conservative force, $\mathbf{F}_{ij}^C$ (eq. S4), controlled by the magnitude of the parameter, $a_{ij}$. This parameter is most commonly set using the method introduced by Groot and Warren [1].

$$\mathbf{F}_{ij}^C = a_{ij} \omega_{ij}(\mathbf{r}_{ij}) \mathbf{e}_{ij} \tag{S4}$$

All forces in eq. S2-S4 are calculated via a weight function that begins at unity and vanishes at the cut-off distance:

$$\omega_{ij}(\mathbf{r}_{ij}) = \begin{cases} \left(1 - \dfrac{r_{ij}}{r_c}\right); & r_{ij} \leq r_c \\ 0; & r_{ij} \geq r_c \end{cases} \quad (S5)$$

Since DPD is a coarse-grained method, the interactions between individual DPD particles are soft, meaning that particles only interact through different forces if they overlap with one another [1].

Particle positions and velocities in each time step are advanced based on a velocity-Verlet algorithm (eq. S7-S10). In this scheme, predictions of particle position (eq. S7) and velocity (eq. S8) are made based on the total force acting on a particle ($\mathbf{F}_i(t)$), particle velocity ($\mathbf{v}_i(t)$), and positions ($\mathbf{r}_i(t)$) in the previous step:

$$\mathbf{r}_i(t + \Delta t) = \mathbf{r}_i(t) + \Delta t \mathbf{v}_i(t) + \frac{1}{2}(\Delta t)^2 \mathbf{F}_i(t) \quad (S7)$$

$$\tilde{\mathbf{v}}_i(t + \Delta t) = \mathbf{v}_i(t) + \frac{1}{2}\Delta t \mathbf{F}_i(t) \quad (S8)$$

In the second step, new forces are calculated through eq. S2-S4, and S6, and corrections are made to the particle velocity (eq. S10) based on this new total force (eq. S9).

$$\mathbf{F}_i(t + \Delta t) = \mathbf{F}_i(\mathbf{r}(t + \Delta t), \tilde{\mathbf{v}}(t + \Delta t)) \quad (S9)$$

$$\mathbf{v}_i(t + \Delta t) = \mathbf{v}_i(t) + \frac{1}{2}\Delta t (\mathbf{F}_i(t) + \mathbf{F}_i(t + \Delta t)) \quad (S10)$$

Throughout the manuscript quantities are expressed in non-dimensional form. To do this, one should choose an energy scale, $kT$, a length scale which is commonly set by the cut-off distance, $r_c$, and the mass of individual particles, $m$, and derive the units for other quantities. Subsequently one will have the following units: $(kT/m)^{1/2}$ for velocity, $r_c(m/kT)^{1/2}$ for time, $kT/r_c$ for force, $kT/mr_c$ for acceleration, $(kT/m)^{1/2}/r_c$ for shear rate and $k_B T/r_c^3$ for stress/pressure/modulus.

Further details of the simulation scheme, parametrization and measurement of rheological properties can be found in number of previous reports [4-7].

**Stress autocorrelation function**

The rescaled stress autocorrelation function, $C_{\alpha\beta}(t) = V\langle \sigma_{\alpha\beta}(t)\sigma_{\alpha\beta}(t+\Delta t)\rangle / kT$, based on the long-time tail of the decay function, referred to as residual stress in the manuscript, can be fitted to an exponential decay function, with a single relaxation time (eq. S11):

$$C_{\alpha\beta}^{fit}(t) = C(0)\exp\left[-t/\tau^*\right] + \sigma_r \tag{S11}$$

In addition to $\sigma_r$, which is used as a proxy measure of the yield stress of the resulting fluids at the end of the Time-Rate-Transformation (TRT) protocol, values of $C(0)$ and $\tau^*$ are presented in Table S1. Subsequently, one can calculate the period of oscillations observed in figure 2 of the manuscript, by fitting the residual value of the fitted data in Fig. S1, $\varepsilon(t) = C_{\alpha\beta}(t) - C_{\alpha\beta}^{fit}(t)$ to a simple sinusoidal function of the form $\varepsilon(t) = \sin\left(\pi\, t/\lambda\right)$. Fig. S1 shows the value of the rescaled stress autocorrelation function after subtracting the residual stress, and the residual value of fitted data as a function of time, for the same systems presented in fig. 2 of the main manuscript. It should be noted that since no visible oscillation is observed for the quench time of $t_q = 500\tau$, no period of oscillation is calculated for that curve in table S1. Note that here $\tau$ is the Brownian diffusive time equal to $a^2/D = 6\pi\eta a^3/kT$ with $D$ the diffusion coefficient of a single particle.

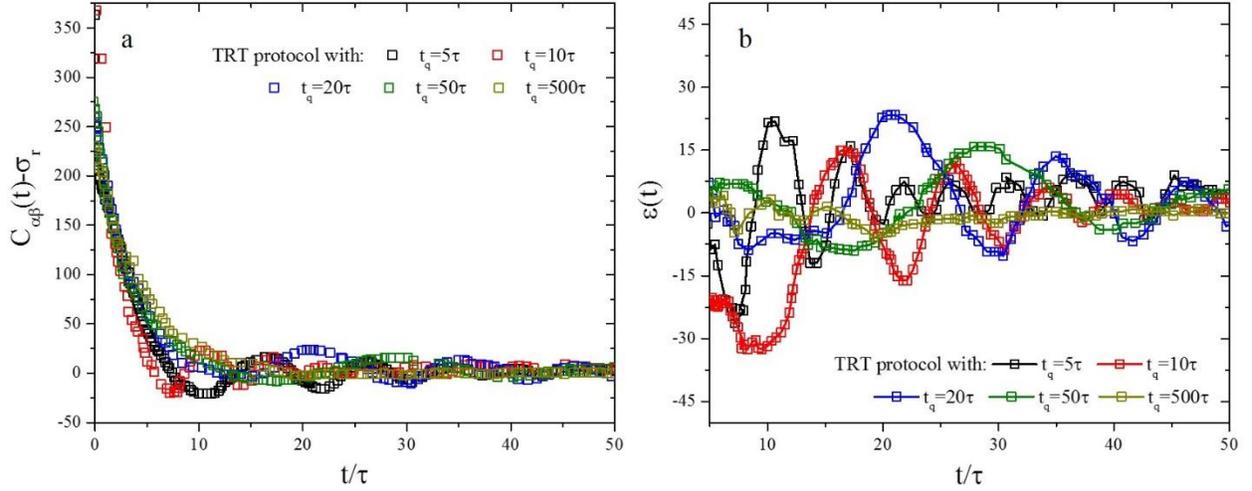

Figure S1. a) Rescaled stress autocorrelation function without residual stress values, and b) fitting residuals as a function of time, for attractive suspensions with $\kappa^{-1} = 0.1a$ and $\Gamma = 30kT$ quenched from $Mn = 1$ over varying quench times.

Table 1. Values of $C(0), \tau^*$ and $\lambda$ calculated from equation S1 and sinusoidal fit.

| Quench Time | $C(0)$ [$kT/a^3$] | $\tau^*/\tau$ | $\lambda/\tau$ |
|---|---|---|---|
| 5 | 375.53437 | 1.71989 | 3.45017 |
| 10 | 366.28178 | 2.88849 | 5.05115 |
| 20 | 360.41218 | 3.05321 | 8.75903 |
| 50 | 261.26967 | 3.17601 | 12.76347 |
| 500 | 210.07016 | 4.51079 | N/A |

As evident in fig. S2, the period of oscillations calculated for quenched suspensions decreases by increasing the attraction strength between the particles. This can be attributed to stronger gels formed at higher attraction strengths, hence larger yield stress and enhanced nonlinear viscoelasticity of those systems; however, the initial amplitude of the stress autocorrelation function is directly related to the relaxation modulus of any given structure. This is more evident when the Green-Kubo [8, 9] expression is considered ($\eta = \dfrac{V}{10kT} \int_0^\infty \langle \sigma_{\alpha\beta}(t)\sigma_{\alpha\beta}(t+\Delta t)\rangle dt$).

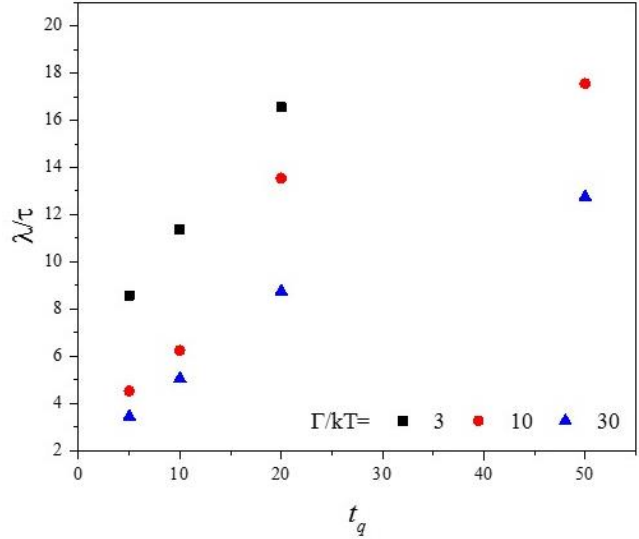

Figure S2. Period of oscillation ($\lambda/\tau$) as a function of quench time for three different strengths of attraction for attractive suspensions with $\kappa^{-1} = 0.1a$ quenched from $Mn = 1$.

## Distribution of coordination number

Fig. S3 shows the probability of finding particles with given coordination numbers for four different accumulated strains, compared to the same probability measured for the gel formed under quiescent conditions for a suspension with $\Gamma = 10kT$ and $\kappa^{-1} = 0.1a$. At fast quench rates, corresponding to small accumulated strains $\gamma_n \leq 10$, the number of particles with large coordination numbers is smaller than observed under equilibrium conditions, this fraction progressively increases with strain, suggesting that particle-rich regions are being formed as the total shear strain experienced by the system accumulates.

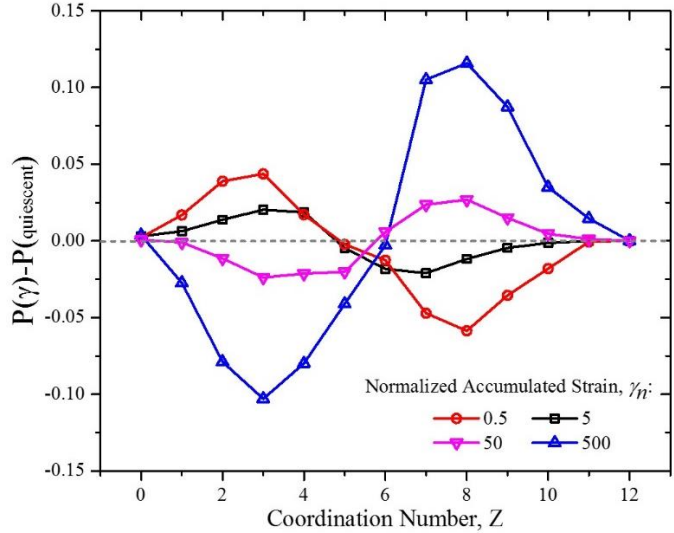

Figure S3. Distribution of coordination numbers for gels at different normalized accumulated strains compared to their values at quiescent conditions. The dashed line represents the value corresponding to the gel formed at rest.

**Material state for different particle characteristics**

The final states of the colloidal systems studied in the manuscript (denoted by the filled squares (■) and shaded with orange color background for "strong gel", hollow circles (□) shaded with yellow color background for "weak gel", and indicated by (×) and shaded with a green background color for "flocculated fluid" states ) are presented in fig. S4. The top figure shows a system of attractive colloids with volume fraction of $\phi = 0.15$ and attraction strength of $\Gamma = 10kT$, brought to rest through our Time-Rate Transformation (TRT) protocol, with varying total accumulated strains, $\gamma_{max}$, and attraction ranges, $\kappa^{-1}$. It is clear from the upper figure that the boundaries of the different final states vary systematically with the range of attraction between the particles when plotted against the total accumulated strain $\gamma_{max}$ imposed during the quench. This motivates the definition of the normalized strain, $\gamma_n$, which reflects the differing ranges of attraction. The bottom figure shows the entirety of our simulation results for a number of different attractions strengths and attraction ranges, versus this normalized accumulated strain, $\gamma_n$. Close inspection of Fig S4 shows that the boundaries between the different gel states that we identify still vary with the strength of the attraction between the particles. To account for this we need to recognize that the thixotropic time scale of these particulate gels also changes systematically with the colloidal attraction strength. Varying quench times thus correspond to different relative timescales for structural evolution in systems with different attraction strengths. In the next section we describe how we determine the characteristic thixotropic time scales for each gel and these values are presented in Fig S5. Once this variation in material time-scale is determined, the effective total time of shearing (weighted by the appropriate time scale for thixotropy in system) is reflected by the product $\gamma_n . \tau_{thix.} / \tau$. When the phase boundaries are plotted in terms of this weighted strain parameter the boundaries between the phases shown in Fig S4 become vertical lines, as shown in the lower portion of Fig 4 of the main manuscript.

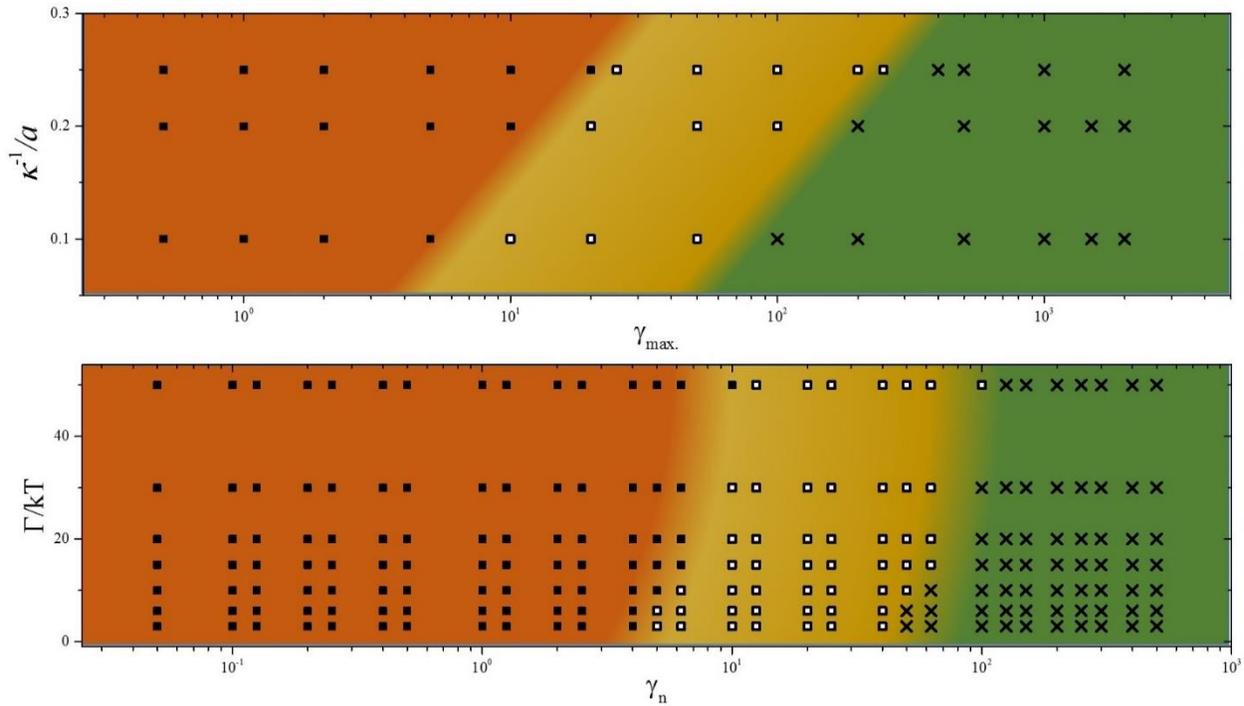

Figure S4. Final material states (based on the criteria defined in the manuscript) for attractive colloids of: (top) varying attraction ranges with fixed attraction strength of $\Gamma = 10kT$ against total accumulated strain, and; (bottom) for varying attraction strengths (and all attraction ranges) against the rescaled accumulated strain,. The final states shown in the lower figure are for a fixed volume fraction of $\phi = 0.15$ and three ranges of attractions, $\kappa^{-1} = 0.1a$, $0.2a$ and $0.25a$ (■ shaded in orange color background for strong gel, □ shaded in yellow color background for weak gel, and × shaded in green background color for flocculated fluid states).

**Thixotropic time scale**

The variation in the thixotropic timescale (scaled by the diffusive time of a single Brownian particle) of attractive suspensions with volume fraction of $\phi = 0.15$ and different attraction strengths, is shown in fig. S5. These timescales are calculated by imposing a series of step-wise ramp-down followed by ramp-up shear protocols, as first proposed by Divoux et al. [8] and subsequently used by Radhakrishnan et al. [9] The computational details are explained in a recent study [10]. In brief, we characterize the rheological hysteresis area in the flow curves of stress vs. shear-rate which are computed through a series of step-rate experiments using varying waiting times at each imposed deformation rate. The resulting hysteresis area varies as a log-normal distribution of the waiting time, with the maximum area located at the characteristic thixotropic time scale, $\tau_{thix.}$. A power law scaling of the data in fig. S5 provides a functional form for calculation of thixotropic timescales as a function of attraction strength between the particles:

$$\tau_{thix.}(\Gamma)/\tau = A(\Gamma/kT)^B \tag{S12}$$

Where $A = 12.93$, $B = -0.32$ with $R^2 = 0.966$ for the fitting confidence. For very strong attraction strengths and short thixotropic timescales the step rate protocol prevents sampling the full log-normal distribution and results in the small deviation observed for $\Gamma = 50kT$. It should be noted that the values of the thixotropic time scale will vary with the range of attraction as well as with the volume fraction of particles; by increasing the effective volume fraction of colloids in the matrix the thixotropic timescale will decrease due to higher probabilities of particle-particle interactions.

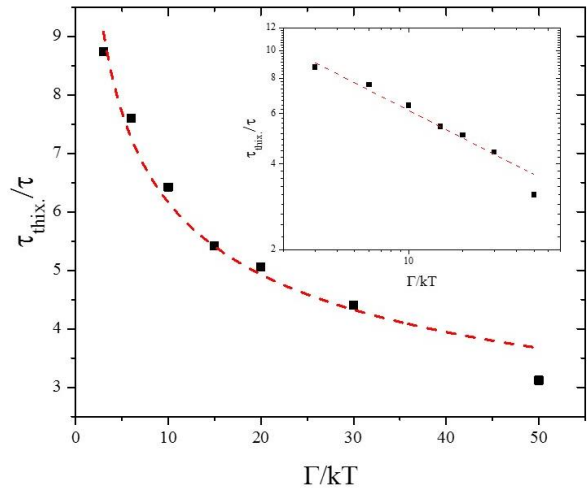

Figure S5. Thixotropic time scale normalized with the diffusive time scale of an individual Brownian particles, against the strength of attraction between two particles, as determined based on rheological hysteresis computations as described in detail in [10]. The red dashed-line presents fitting of the data to an exponential decay function of the form $\tau_{thix.}/\tau(\Gamma) = A(\Gamma/kT)^B$. The inset shows the same data in a log-log plot.

**Alternative flow decay protocols for TRT**

Three different types of flow decay protocols are compared in fig. S6 based on the temporal variation in the applied shear rate (or Mason number $Mn(t)$) and accumulated strains, corresponding to three different quench times: $t_q/\tau$ = 1, 10 and 100. The kinematics of the flow for the linear, exponential and step-wise decay functions can be written as eq. S13-S15, respectively:

$$\dot{\gamma}(t) = \dot{\gamma}_0 \left(1 - \frac{t}{t_q}\right) \quad \text{(S13)}$$

$$\dot{\gamma}(t) = \dot{\gamma}_0 \exp(-t/t_q) \quad \text{(S14)}$$

$$\dot{\gamma}(t) = \dot{\gamma}_0 \left(1 - \text{int}\{Nt/t_q\}/N\right) \quad \text{(S15)}$$

Where $N$ in equation S15 is the number of steps to be taken towards zero shear rate, and the int[10] operator returns the integer value of the argument $Nt/t_q$.

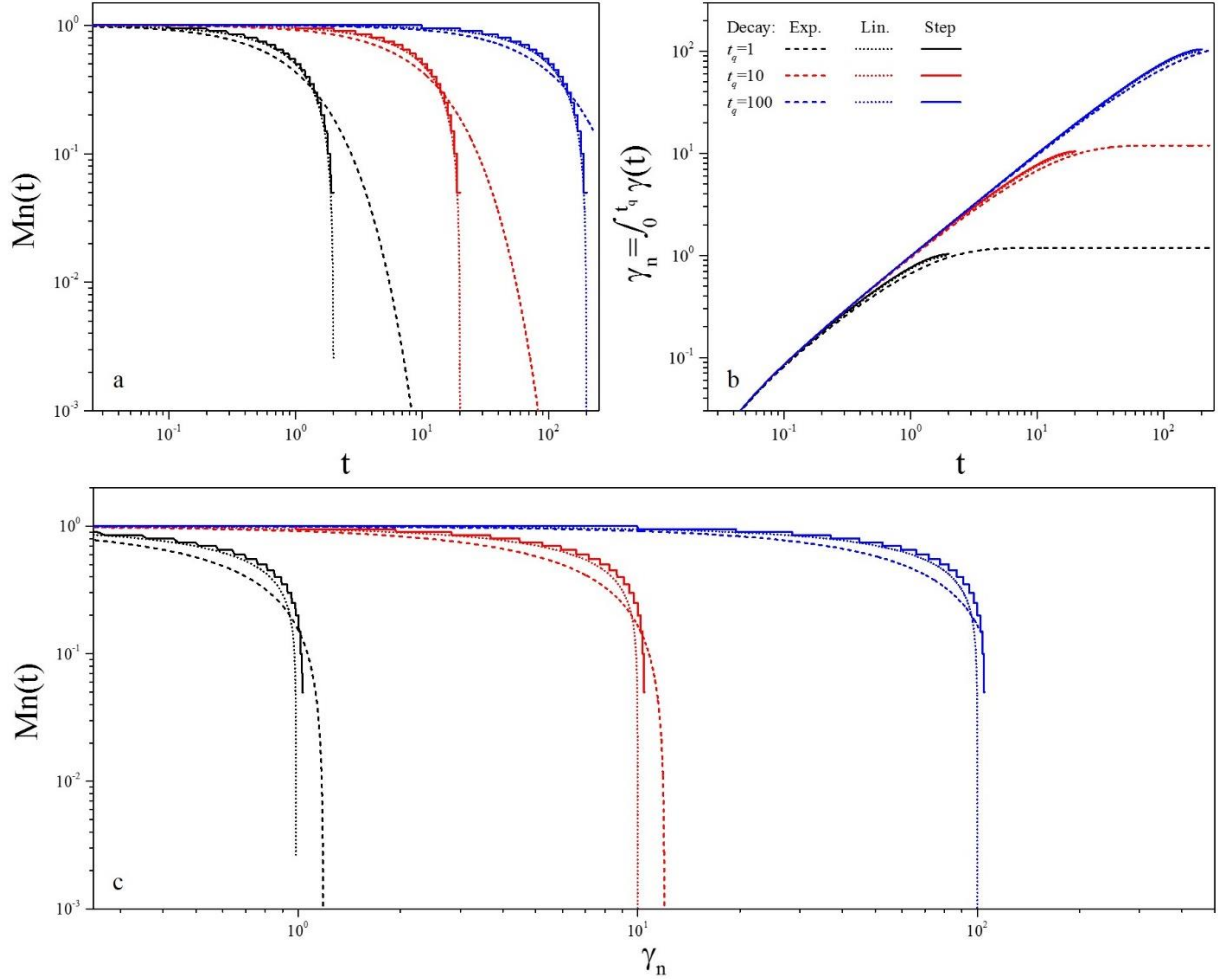

Figure S6. Time-varying Mason number (a), normalized accumulated strain as a function of time (b), and (c) Mason number as a function of normalized accumulated strain for three different flow decay protocols and three different quench times, corresponding to the strong gel, weak gel, and flocculated fluid regions in our TRT phase diagram in figure 4 of the main manuscript.

We compute the final states of the resulting gels obtained for each of the trajectories shown in Figure S6 using the three distinct structural measures described in the main text (see discussion related to Figs 2 & 3). Table S2 shows the three measurables defined in the manuscript (rescaled residual stress, rescaled coordination number and rescaled number density fluctuations), as well as the final material state for attractive colloidal systems with $\Gamma = 10kT$ and $\kappa^{-1} = 0.2a$ quenched with three different protocols described through equations S13-S15.

| Decay Function | $t_q$ | $\sigma_r$ | $\mathcal{Z}$ | $\mathcal{N}$ | Material State |
|---|---|---|---|---|---|
| Linear (eq S13) | 1 | 1.01 | 0.94 | 0.99 | Strong Gel |
| | 10 | 0.48 | 0.97 | 2.12 | Weak Gel |
| | 100 | 0.03 | 1.06 | 3.67 | Flocculated Fluid |
| Step-wise (eq S14) | 1 | 0.99 | 0.97 | 1.01 | Strong Gel |
| | 10 | 0.37 | 1.00 | 2.30 | Weak Gel |
| | 100 | 0.01 | 1.12 | 3.86 | Flocculated Fluid |
| Exponential (eq S15) | 1 | 1.02 | 0.92 | 0.97 | Strong Gel |
| | 10 | 0.41 | 0.99 | 1.96 | Weak Gel |
| | 100 | 0.01 | 1.09 | 4.08 | Flocculated Fluid |

Despite the small numerical differences in the exact of $\mathcal{Z}$, $\mathcal{N}$ and $\sigma_r$ shown in Table S2, it is clear that the same distinct gel states are obtained for each different quench protocol with the same quench time-scale. It is thus clear that the Time-Rate-Transformation (or Strain-Mason Number transformation) framework we outline in this work is useful for systematically understanding the role of shear flow history and targeting specific colloidal microstructures in these thixotropic elastoviscoplastic systems.

**Role of interaction potential**

Investigation of gel structure and properties with respect to Mason number [being solely connected to the maximum strength of attraction potential] is only justified for short-range attractive particles [11]. As the interaction becomes longer-range, the density of interaction along the particle-particle separation distance affects the final microstructure as well as the mechanical properties of the gel. On the other hand, for short-range attractions, the specific choice of interaction potential becomes rather irrelevant, as one can consider particles within those distances as effectively in contact with one another [10]. In order to ensure that our choice of interaction type (i.e. the soft Morse potential) does not influence the universality of our conclusions, we have also tested two other commonly used interaction potentials for a number of attractive gels under Time-Rate-Transformation.

$$U_{AO} = -\Gamma \frac{2(2a(1+\delta))^3 - 3r(2a(1+\delta))^2 + r^3}{2(2a(1+\delta))^3 - 6a(2a(1+\delta))^2 + (2a)^3} \quad \text{(S16)}$$

$$U_{SW} = -\Gamma; 2a \leq r \leq 2(a+\delta) \tag{S17}$$

Equations S16 and S17 respectively show the Asakura-Oosawa potential commonly used for depletion gels and square well potential assuming a constant attraction across the range of attraction ($\delta$). In both equations $r$ is the center-center distance of two interacting particles, and $a$ is the radius of a particle. One should note that the Morse potential used in this study is cut off at the attraction range ($\kappa^{-1}/a =$ 0.1, 0.2 and 0.25) as well and thus is not a long-range attraction. Figure S7 shows the typical corresponding interaction potentials with values of $\kappa^{-1}/a = 0.1$ and $\delta = 0.1$ for AO, Morse and square-well potentials.

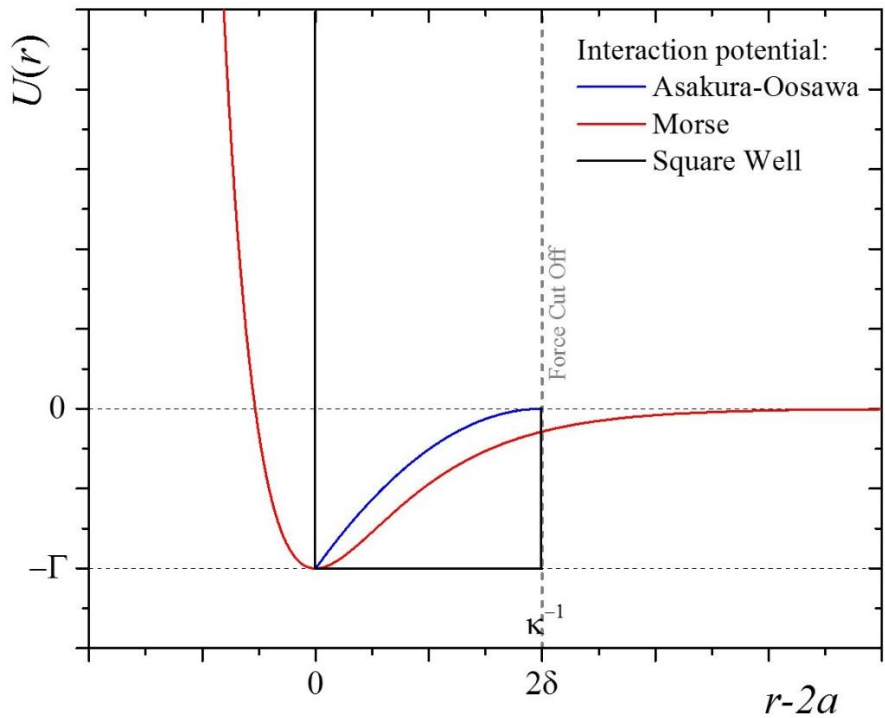

Figure S7. The interaction potentials of different types (Asakura-Oosawa, Morse and Square Well) for a given range of attraction, and a corresponding value of strength.

Using the functional forms of each interaction potential, one can calculated the stickiness parameter, $\tau$ (different from diffusion time in the manuscript), and the reduced second virial coefficient of each interaction type, using the formalism of Noro and Frenkel [10]. Table S3 shows the measured values of these parameters for each interaction potential, for a given value of the attraction strength ($10kT$), and cut off range ($0.25a$). Since the values of second virial coefficients are similar for all three interaction potentials, having the same interaction strength and cut off, the phase diagram of the resulting structures for all interaction potentials are expected to be similar. This further proves that in the short-range regime we have investigated the choice of potential shape does not affect the final structures.

| **Interaction type** | $\tau$ | $B_2^*$ |
|---|---|---|
| Morse | 0.103 | 0.975 |
| Asakura-Oosawa | 0.112 | 0.972 |
| Square Well | 0.099 | 0.976 |

Table S4 shows the three measurables defined in the present manuscript (rescaled residual stress, rescaled coordination number and rescaled number density fluctuations), as well as the final material state for attractive colloidal systems with $\Gamma = 10kT$ and $\kappa^{-1}/a = 0.1$ quenched with three different protocols for particles interacting through equations S16 and S17 as well as with the Morse potential used in the rest of this work.

| Decay Function | $t_q$ | $\sigma_r$ | $\mathcal{Z}$ | $\mathcal{N}$ | Material State |
|---|---|---|---|---|---|
| Morse | 1 | 1.01 | 0.94 | 0.99 | Strong Gel |
| | 10 | 0.48 | 0.97 | 2.12 | Weak Gel |
| | 100 | 0.03 | 1.06 | 3.67 | Flocculated Fluid |
| Asakura-Oosawa | 1 | 0.99 | 0.94 | 0.98 | Strong Gel |
| | 10 | 0.46 | 0.96 | 2.18 | Weak Gel |
| | 100 | 0.02 | 1.08 | 3.72 | Flocculated Fluid |
| Square Well | 1 | 1.01 | 0.95 | 0.97 | Strong Gel |
| | 10 | 0.49 | 0.97 | 2.10 | Weak Gel |
| | 100 | 0.04 | 1.05 | 3.59 | Flocculated Fluid |

Results presented in Tables S2 and S3 indicate that once the attraction between particles is restricted to the same range of interaction [and remain within the short-range limit], the choice of interaction potential does not change the final results and our quantitative measures become insensitive to such potential energy distribution.

# References

1. Groot, R.D. and P.B. Warren, *Dissipative particle dynamics: Bridging the gap between atomistic and mesoscopic simulation.* Journal of Chemical Physics, 1997. **107**(11): p. 4423-4435.
2. Espanol, P., *Hydrodynamics from dissipative particle dynamics.* Physical Review E, 1995. **52**(2): p. 1734-1742.
3. Espanol, P. and P. Warren, *Statistical Mechanics of Dissipative Particle Dynamics.* Europhysics Letters, 1995. **30**(4): p. 191-196.
4. Boromand, A., S. Jamali, and J.M. Maia, *Viscosity measurement techniques in Dissipative Particle Dynamics.* Computer Physics Communications, 2015. **196**: p. 149-160.
5. Boromand, A., S. Jamali, and J.M. Maia, *Structural fingerprints of yielding mechanisms in attractive colloidal gels.* Soft Matter, 2017. **13**(2): p. 458-473.
6. Jamali, S., et al., *Gaussian-inspired auxiliary non-equilibrium thermostat (GIANT) for Dissipative Particle Dynamics simulations.* Computer Physics Communications, 2015. **197**: p. 27-34.
7. Jamali, S., G.H. McKinley, and R.C. Armstrong, *Microstructural rearrangements and their rheological implications in a model Thixotropic Elastoviscoplastic fluid.* Physical Review Letters, 2017. **118**(4): p. 048003.
8. Kubo, R., *Statistical-mechanical theory of irreversible processes. I. General theory and simple applications to magnetic and conduction problems.* Journal of the Physical Societ of Japan, 1957. **12**(6): p. 570-586.
9. Lin, N.Y.C., M. Bierbaum, and I. Cohen, *Determining quiescent colloidal suspension viscosities using the Green-Kubo relation and image-based stress measurements.* Physical Review Letters, 2017. **119**(13): p. 138001.
10. Noro, M.G. and D. Frenkel, *Extended corresponding-states behavior for particles with variable range attractions.* The Journal of Chemical Physics, 2000. **113**(8): p. 2941-2944.
11. Lekkerkerker, H.N. and R. Tuinier, *Colloids and the depletion interaction.* Vol. 833. 2011: Springer.